\documentclass[twocolumn,superscriptaddress,aps,longbibliography]{revtex4-1}
\usepackage{natbib}
\usepackage{kotex}
\usepackage{graphicx}
\usepackage{enumitem}
\DeclareGraphicsExtensions{.pdf,.png,.jpg}
\usepackage{array} 
\usepackage{amsmath} 
\usepackage{amssymb}    
\usepackage{latexsym}
\usepackage{amsfonts}
\usepackage{mathrsfs}
\usepackage{color}
\usepackage{bbold}
\usepackage{gensymb}
\usepackage{siunitx}
\usepackage{verbatim} 

\usepackage{subfigure}
\usepackage[colorlinks=true, urlcolor=blue,citecolor=blue,linkcolor=blue]{hyperref}

\begin{document}
\newcommand{\bra}[1]{\left< #1\right|}   
\newcommand{\ket}[1]{\left|#1\right>}
\newcommand{\abs}[1]{\left|#1\right|}
\newcommand{\avg}[1]{\left<#1\right>}
\newcommand{\Tr}{\mbox{Tr}}
\renewcommand{\d}[1]{\ensuremath{\operatorname{d}\!{#1}}}

\title{Optical experiment to test negative probability in context of quantum-measurement selection}
\author{Junghee Ryu}
\email{rjhui82@gmail.com}
\affiliation{Centre for Quantum Technologies, National University of Singapore, 3 Science Drive 2, 117543 Singapore, Singapore}

\author{Sunghyuk Hong}
\author{Joong-Sung Lee}
\author{Kang Hee Seol}
\author{Jeongwoo Jae}
\affiliation{Department of Physics, Hanyang University, Seoul 04763, Republic of Korea}

\author{James Lim}
\affiliation{Institute of Theoretical Physics and Integrated Quantum Science and Technology IQST, University of Ulm, Albert-Einstein-Allee 11, D-89069 Ulm, Germany}

\author{Jiwon Lee}
\author{Kwang-Geol Lee}
\email{kglee@hanyang.ac.kr}
\author{Jinhyoung Lee}
\email{hyoung@hanyang.ac.kr}
\affiliation{Department of Physics, Hanyang University, Seoul 04763, Republic of Korea}

\begin{abstract}
Negative probability values have been widely employed as an indicator of the nonclassicality of quantum systems. Known as a quasiprobability distribution, they are regarded as a useful tool that provides significant insight into the underlying fundamentals of quantum theory when compared to the classical statistics. However, in this approach, an operational interpretation of these negative values with respect to the definition of probability---the relative frequency of occurred event---is missing. An alternative approach is therefore considered where the quasiprobability operationally reveals the negativity of measured quantities. We here present an experimental realization of the operational quasiprobability, which consists of sequential measurements in time. To this end, we implement two sets of polarization measurements of single photons. We find that the measured negativity can be interpreted in the context of selecting measurements, and it reflects the nonclassical nature of photons. Our results suggest a new operational way to unravel the nonclassicality of photons in the context of measurement selection.
\end{abstract}

\maketitle



{\it Introduction.}---As previously discussed by Richard P. Feynman, a negative probability, which relaxes the axiom of a non-negative probability of an event in Kolmogorov's probability theory, sheds new light on our understanding of quantum phenomena~\cite{Feynman, Feynman82}. 
The essence of his idea is that a negative probability results in much less mathematical complications in intermediate steps for the analysis of a given physical event. As an example, Feynman developed joint probability distributions for spin-$1/2$ systems to address Young's double-slit experiment using a different approach, such that the probability distributions can have negative values~\cite{Feynman}. Such an idea has been applied to many studies involving various quantum phenomena~\cite{Scully94, Home91, Higgins15, Spekkens08, Han96}. In particular, the negative probability approach provides new insight into a disagreement between classical and quantum predictions on the Bell's theorem~\cite{Bell64}. It is experimentally verified that our nature cannot be described by any classical theory of local realism, as the classical theory obeys but the quantum theory violates the Bell inequalities~\cite{Gus, hen, Shalm}. Negative probability introduces a different point of view that local realistic theory endowed with a negative probability can simulate violations of the Bell inequalities, similar to quantum theory~\cite{Sudarshan93, Cereceda00, Rothman01}. However, these negative probabilities are distinct from the definition of probability with respect to the relative frequency of events, in which case, an operational interpretation is not necessarily straightforward (for historical review, see Ref.~\cite{Muckenheim86}).


On the other hand, negative values play a role as an indicator. 
Let us retrace Feynman's example on the trading of apples~\cite{Feynman}: ``A man starting a day with five apples who gives away ten and is given eight during the day has three left.'' Feynman pointed out that the initial and final numbers of apples, we denote by $(5,3)$ and call a fundamental entity, can be calculated in the following two ways. The first case reads ``beginning with 5 apples at time $t_1$, given 8 at $t_2$, giving away 10 at $t_3$, and having 3 left at $t_4$'', or symbolically, $5_{t_1}+8_{t_2}-10_{t_3}=3_{t_4}$ with a time ordering $t_i < t_j$ for $i < j$. This can be claimed to be natural since the apple number is always kept as non-negative. On the other hand, Feynman introduced an alternative model which disregards keeping the apple number as nonnegative but maintains the same fundamental entity $(5_{t_1}, 3_{t_4})$; it reads $5-10=-5$, and $-5+8=3$. Note that the latter model disregards the chronological order $t_i$, i.e., changes the order of trading, thus a negative number of apples appears in the intermediate steps.
Although such negative numbers are abstract, allowing negativity leads to have more freedom in mathematical analysis without altering the fundamental entity. We say that such a model is {\em free of time context}~\cite{Spekkens05,Ferrie11}. In other words, the time context-free approach maintains the same fundamental entity at the expense of relaxing the assumption of nonnegative apples. Moreover, the negative apples in the intermediate steps introduce the idea of ``debts'' with respect to the trading of apples (the indicator).


In quantum optics, negative values have been widely used as an indicator of nonclassicality in regard to classical statistics. For instance, the Wigner function, one of the so-called quasiprobability distributions, is used to represent a joint distribution of the position and momentum in phase space~\cite{Wigner32}. However, due to the uncertainty principle with such conjugate variables, the Wigner function is negatively valued for some quantum states. It witnesses quantum phenomena of the states and has been generalized to finite dimensional systems such as quantum bits, and currently the quasiprobability approach has been applied to the omnidirectional range of fields in quantum information science~\cite{Leonhardt95, Vaccaro90, Miquel02a,Miquel02b,Paz02,Koniorczyk01,Durt08,Waller12,Howard14,Sperling18,Sperling19a,Sperling19b}.

However, such nonclassicality indicators with Wigner functions lack the operational formalism wherein preparation, operation, and measurement cooperate explicitly~\cite{Spekkens05,Ferrie11}. Negative values in one quasiprobability can be positive in another. This fact can be an obstacle to the operational interpretation of these negative values. Furthermore, their comparison to classical statistics reveals the subtlety that quasiprobabilities require different physical interpretations for the same form of functionals as their classical counterparts. This is described as ``incommensurability'' of quasiprobabilities~\cite{Ryu13, Jae17}. Therefore, it is more natural to employ quasiprobability that compares quantum and classical statistics on the same footing.

The operational quasiprobability (OQ) introduced in~\cite{Ryu13, Jae17} allows the problems in question to be resolved. The OQ is commensurate since it evaluates the statistics of the mathematical functionals with the same physical interpretation in every model, regardless of whether a quantum or classical case is being considered. The OQ consists of selective and sequential measurements in time and is formulated as joint probability distributions that simultaneously describe multiple measurement setups. We consider the expectation values of multiple measurement setups as fundamental entities, i.e., the measurable quantities of interest. The OQ method allows the joint distributions to be independent of the measurement setups and to simultaneously describe the multiple-setup outcomes, even though the distributions can be negative. This method can be considered as an alternative way of describing quantum theory that depends on the setups. It allows a direct comparison between quantum and classical statistics and identifies nonclassicality in an operational way.

We are focused on the specific feature that the moments will vary depending on the measurement(s) that are performed. This is called measurement-selection context and is similar to the time context of the Feynman's apple example. The fundamental entity is a set of all moments in the single measurements. For two dichotomic observables, it is $\{(\langle A^n \rangle , \langle B^m \rangle)\}$ with $\langle A^n\rangle$ being $n$-th moments in one of the single measurements and similarly $\langle B^m \rangle$ in the other. The OQ is {\em free of the measurement-selection context}, in the sense that the local marginals of the joint distribution are equal to the probabilities of the single measurements. This is one of the most astonishing features that classical models presume, including a macrorealistic model~\cite{Clemente16}. Moreover, the context-free OQ can always be constructed in quantum theory, even though quantum theory is not context-free. Instead, the quantum OQ pays a tariff of negative probabilities. Such inevitable negatives can be understood as an indicator of nonclassicality in the context of measurement selection.

The macrorealistic model assumes no-signaling in time (NSIT) and arrow of time (AoT); NSIT implies that a later measurement produces a result which is not affected by whether or not an earlier measurement is performed, and AoT is a similar condition where the role of later and earlier measurements is exchanged. In other words, roughly speaking, the two measurements are independent, and both of the sequential measurement and the individual measurements leave the fundamental entity unchanged. The classical model is thus free of the measurement-selection context. More explicitly, when two measurements are sequentially performed at times $t_{1}$ and $t_2$ with $t_1 < t_2$, respectively, or they are individually performed, the NSIT and AoT are described by
\begin{eqnarray}
\text{NSIT :}&&~ P_{t_2}(a_2) =P_{t_1,t_2}(a_2), \nonumber\\ 
\text{AoT :}&&~ P_{t_1}(a_1) = P_{t_1,t_2}(a_1). \nonumber 
\end{eqnarray}
Here $P_{t_i}(a_i)$ are the probabilities of the single measurements, whereas $P_{t_1, t_2}(a_i)$ are the marginals of the joint measurement
$P_{t_1, t_2}(a_1,a_2)$ such that $P_{t_1, t_2}(a_i) =\sum_{a_{j \neq i}} P_{t_1, t_2}(a_1,a_2)$ for $i,j=1,2$.
In contrast, quantum theory violates the macrorealistic assumptions and it is contextual with the measurement selection. 

However, its OQ representation allows the joint distribution to be free of measurement-selection context as in the macrorealistic model, and it pays the quantum tariff of negative probabilities. The OQ representation states that a quasiprobability distribution ${\cal W}$ is operationally defined for both quantum and classical models by~\cite{Ryu13,Jae17}:
\begin{eqnarray}
\mathcal{W}(a_1,a_2) = P_{t_1,t_2}(a_1,a_2) &+& \frac{1}{2}\left[ P_{t_1}(a_1) - P_{t_1,t_2}(a_1) \right] \nonumber\\
&+& \frac{1}{2}\left[ P_{t_2}(a_2) - P_{t_1,t_2}(a_2) \right].
\label{eq:w}
\end{eqnarray}
The ${\cal W}$ has marginal probabilities equal to those of the single measurements. The presence of negative ${\cal W}(a_1, a_2)$ originates from the statistical difference between the single and sequential measurements. Note that ${\cal W}$ will be nonnegative joint probabilities if NSIT and AoT are applicable. Thus, measuring OQ represents an evaluation of the context-free model for a given experiment on one hand, and also is a test of whether a given system violates the macrorealistic model based on its negative values on the other hand. Recently, experiments to test macrorealism have been conducted with various quantum states and measurement schemes~\cite{Laloy10,Xu11,Knee12,Knee16}.

It is worth noting that a negative OQ suffices for the failure of the macrorealistic description but is not necessary. Despite the violation of the assumptions, we can still observe a positive OQ depending on the statistical differences. Such a behaviour was reported by employing general measurements~\cite{Jae17}. This implies that the macrorealism, or the conditions of NSIT and AoT cannot fully capture the classicality reflected by the OQ positivity in the context of measurement-selection.

 \begin{figure}[t]
	\centering
	\includegraphics[width=0.45\textwidth]{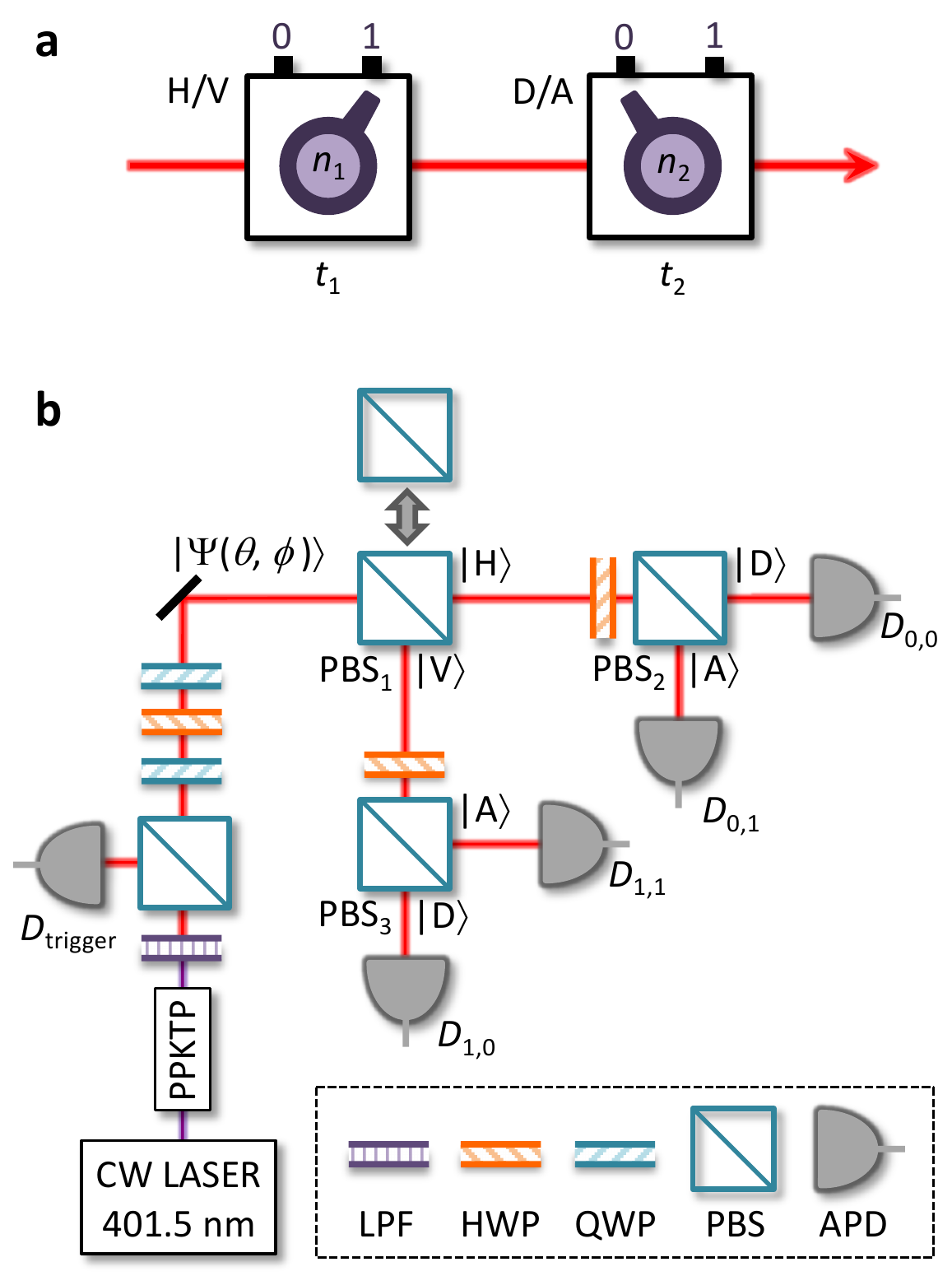}
	\caption{{\bf Experimental scheme for measuring the polarization of single photons.} {\bf a,} A sequential polarization measurement was implemented. Two knobs denoted by $n_1$ and $n_2$ indicate the selective polarization measurements at the times $t_1$ and $t_2$, respectively. Such configurations are implemented in a laboratory setting, as shown {\bf b}. When the PBS$_{1}$ is in position, the H polarization is measured by taking the sum of clicks on $D_{0,0}$ and $D_{0,1}$ and the V polarization is obtained from the sum of $D_{1,0}$ and $D_{1,1}$. If the PBS$_{1}$ is out, then the input light travels directly to the PBS$_{2}$ corresponding to the D/A polarization measurement (see Methods for more details).}
	\label{FIG:exp_scheme}
\end{figure}

\begin{figure*}[t]
	\centering
	\includegraphics[width=0.8\textwidth]{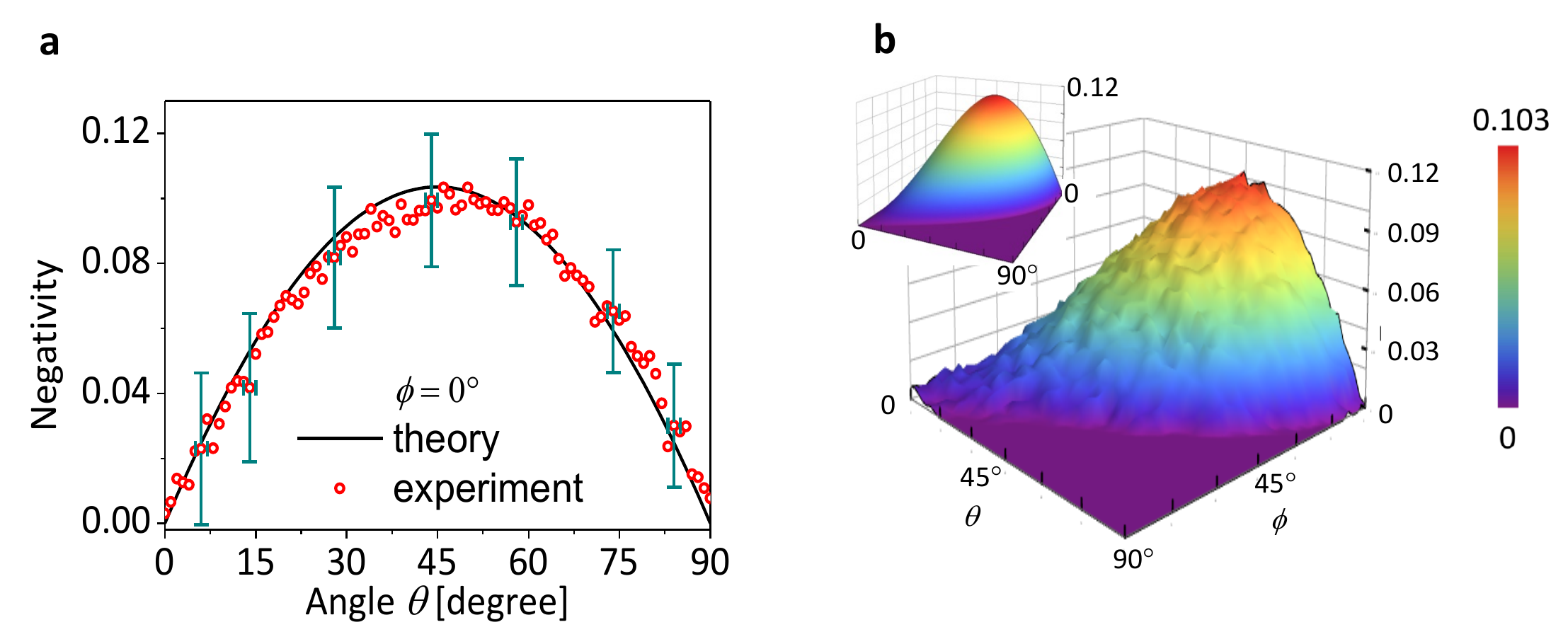}
	\caption{{\bf Negativity with a source by spontaneous parametric down conversion}. {\bf a}, Negativity as a function of $\theta$ with a fixed value of $\phi=0^{\circ}$ of the state $\ket{\Psi ( \theta, \phi)} = \cos (\theta/2) \ket{H} + e^{i\phi} \sin (\theta/2)\ket{V}$. The black line shows the theoretical values and the experimental results are denoted by the red circles with the error estimation. {\bf b}, A contour plot of the measured negativity for $0 \leq \theta, \phi \leq 90^{\circ}$. The inset represents the theoretical values. The experimental maximum negativity $\mathcal{N} \approx 0.103$ is obtained at $\theta=45^{\circ}$ and $\phi=0^{\circ}$.
}
	\label{FIG:neg_spdc}
\end{figure*}

We now experimentally illustrate the negative probability in an operational way by measuring the degrees of freedom of the polarization of single photons. By considering the negative quasiprobability together with the quantum nature of photons, nonclassicality was demonstrated in the laboratory. To this end, we selectively implemented two sets of polarizations at consecutive times; the horizontal/vertical (H/V) measurement at $t_1$ and the diagonal/anti-diagonal (D/A) at $t_2$. Let us denote the selection of measurements by a tuple $(n_1, n_2)$. We can then perform the following four measurement setups: (i) no measurement by $(n_1,n_2)=(0,0)$, (ii) H/V single measurement by $(n_1,n_2)=(1,0)$, (iii) D/A single measurement by $(n_1,n_2)=(0,1)$, and (iv) consecutive joint measurement of H/V and D/A by $(n_1,n_2)=(1,1)$. In this way, each of the probabilities at times $t_i$ in equation~(\ref{eq:w}) is associated with the tuple. We represent the experimental results by the notation of times for convenience.
For example, the joint probabilities $P_{t_1, t_2}(a_1,a_2)$ are equal to $P_{n_1, n_2}(a_1,a_2)$ with $(n_1, n_2)=(1,1)$.
The negativity is defined by the sum of the negative components of $\mathcal{W}(a_1, a_2)$: $\mathcal{N} \equiv \frac{1}{2}\sum_{a_1, a_2} \left[\abs{\mathcal{W}(a_1,a_2)}-\mathcal{W}(a_1,a_2)\right].$
The maximum of the negativity is $(\sqrt{2}-1)/4 \approx 0.104$ in case of two measurements~\cite{Ryu13}. 
We investigated the negativity of OQ for two input sources: (i) heralded single photons generated by spontaneous parametric down-conversion (SPDC) and (ii) single photons emitted from a single molecule. Note that all inputs were set to generate a single photon.


For the generation of heralded single photons, we exploited collinear type-II SPDC as shown in Fig.~\ref{FIG:exp_scheme}b. The signal was counted only when the trigger ($D_{\text{trigger}}$) clicked. The input polarization state of a single photon is given by $\ket{\Psi ( \theta, \phi)} = \cos (\theta/2) \ket{H} + e^{i\phi} \sin (\theta/2)\ket{V}$ with $\theta, \phi \in [0,90^{\circ}]$, where the angles of waveplates determine $\theta$ and $\phi$ (see Methods). To experimentally implement the four measurement setups for OQ, which are described in Fig.~\ref{FIG:exp_scheme}a, we used the arrangement depicted in Fig.~\ref{FIG:exp_scheme}b. Three polarizing beam splitters (PBSs) and two half-wave plates (HWPs) are used to selectively measure the polarization states of photons. The PBS$_1$ is used for the H/V polarization measurements and the two PBS$_2$ and PBS$_3$ with the HWPs are used for the D/A polarization measurements. Selective measurements can be performed by moving each PBS in and out of the path of the input beam. In the laboratory, we positioned two PBS$_2$ and PBS$_3$ at fixed positions to reduce experimental errors, and moved only the PBS$_1$ to implement the four measurement setups. In the detection part, we counted the relative ratio of measurement outcomes as follows: $P_{n_1,n_2}(a_1,a_2)=N_{n_1,n_2}(a_1,a_2)/N^{\textrm{tot}}_{n_1,n_2}$, where $N_{n_1,n_2}(a_1,a_2)$ denotes the sum of the counted photons at the detector $D_{a_1,a_2}$ and $N^{\textrm{tot}}_{n_1,n_2}$ is the total number of counted photons at all detectors for a given setup $(n_1,n_2)$. See Methods for more details.

Figure~\ref{FIG:neg_spdc} shows the negativity of OQ by the heralded single photon as a function of $\theta$ and $\phi$ in the state $\ket{\Psi ( \theta, \phi)}$. For $\phi=0^{\circ}$, we observed the negativity for all range of $\theta$ (see Fig.~\ref{FIG:neg_spdc}a). The error bars are obtained by considering the functioning errors of the optics and devices used (see Supplementary Information). Figure~\ref{FIG:neg_spdc}b shows a contour plot of the negativity as varying the variables $\theta$ and $\phi$. The maximum negativity yields $\mathcal{N} \approx 0.103$ at $\theta = 45 \degree$ and $\phi= 0\degree$, which well reproduces the theoretical prediction~\cite{Ryu13}. The negativity clearly indicates that any classical models that assume the NSIT and AoT conditions, cannot describe the nonclassicality of a single photon.

\begin{figure*}[t]
	\centering
	\includegraphics[width=1.0\textwidth]{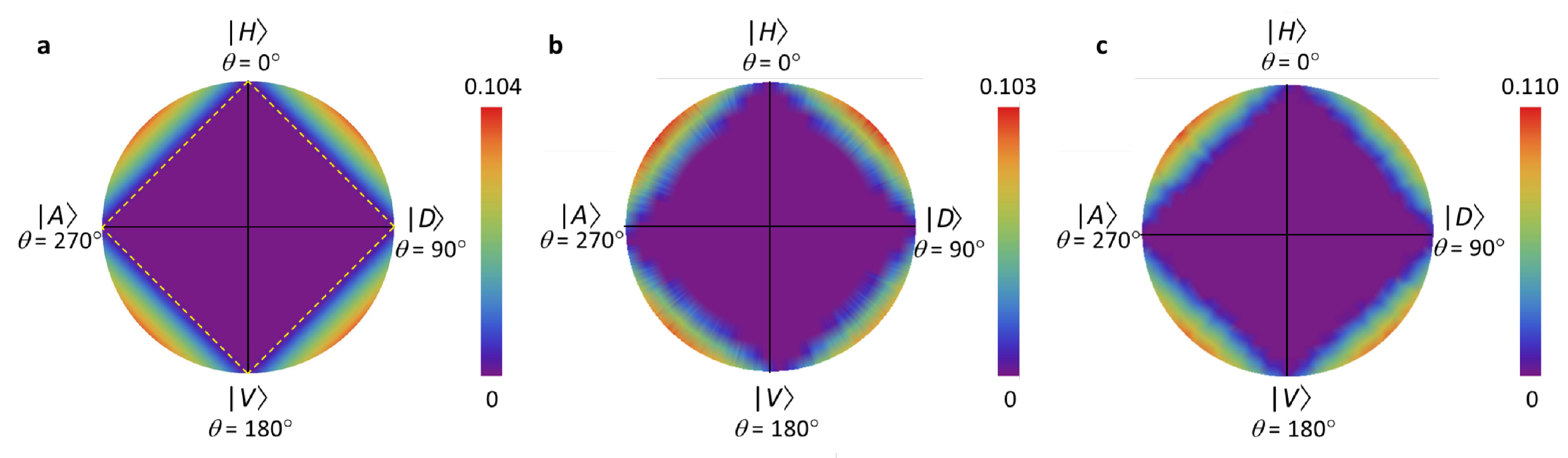}
	\caption{{\bf Negativity of the mixed states}. {\bf a,} Simulation of the negativity in a cross-sectional plane of a Bloch sphere for $\phi=0^{\circ}$. The points on the surface of the circle corresponds to the pure state as $\ket{\Psi ( \theta, \phi=0)} = \cos (\theta/2) \ket{H} + \sin (\theta/2)\ket{V}$. The inner points correspond to the mixed states which can be represented by a probabilistic mixture of the pure states as $\hat{\varrho} (\theta_1, \theta_2, \alpha) = \frac{1+\alpha}{2} \ket{\Psi (\theta_1)} \bra{\Psi (\theta_1)} + \frac{1-\alpha}{2} \ket{\Psi (\theta_2)} \bra{\Psi (\theta_2)}$ with a mixing parameter $\alpha$. The yellow dotted lines denote the mixture of two pure states among the four measuring bases which results in zero negativity for each case. All points inside the dotted square, therefore, give a zero negativity. The centre position corresponds to a completely depolarized state. {\bf b}, Experimental results for the heralded single photons and {\bf c} for single photons from a molecule. We plotted the data for the different conditions of the parameters as $0 \leq \theta_1 \leq 180^{\circ}$, $\theta_2 = \theta_1 + 180^{\circ}$, and $\abs{\alpha} \leq 1$ with resolution values of $\Delta \theta_1 = 1^{\circ}$  and $\Delta \alpha=1/30$. The slightly distorted shape is due to the imperfect alingment and response of the wave-plate in the experiments.}
	\label{FIG:neg_mixed}
\end{figure*}

The quantum nature and the negativity of a single photon is reduced due to the decoherence. Figure~\ref{FIG:neg_mixed}a shows the theoretical values of the negativity on a cross-section of the Bloch sphere for $\phi = 0^{\circ}$. A pure state in the form of $\ket{\Psi ( \theta, \phi=0)} = \cos (\theta/2) \ket{H} + \sin (\theta/2)\ket{V}$ with $ 0 \leq \theta \leq 360^{\circ}$ is placed on the rim of the plane. The inner points correspond to mixed states which are in the probabilistic mixture of the pure states;  $\hat{\varrho} (\theta_1, \theta_2, \alpha) = \frac{1+\alpha}{2} \ket{\Psi (\theta_1)} \bra{\Psi (\theta_1)} + \frac{1-\alpha}{2} \ket{\Psi (\theta_2)} \bra{\Psi (\theta_2)}$, where $\alpha$ determines the mixing ratio of two pure states with a constraint of $\abs{\alpha} \leq 1$. When $\abs{\alpha}=1$, $\hat{\varrho}$ becomes the pure state. The central point of the circle indicates a completely depolarized state. The diamond dash lines in Fig.~\ref{FIG:neg_mixed}a represent the mixture of two pure states, where the negativity becomes zero and thus their convex region inside the dash lines also has a zero negativity. In this case, the experimental results, Fig.~\ref{FIG:neg_mixed}b, are again in well agreement with the theoretical predictions.

The heralded single photons exhibit an anti-bunching feature (see Supplementary Information). However, discussions on the second-order correlation function of these photons have been previously reported~\cite{PFAB04, RAZAVI09, Bashkansky14}. As a demonstration with a deterministic single photon source, we performed similar measurements with photons emitted from a single molecule (terrylene)~\cite{Treussart01} (see Methods for more experimental details). In this case, the photon statistics clearly show the anti-bunching nature without any detection schemes such as triggering (see Supplementary Information). A similar negativity is obtained compared to the SPDC case (see Fig.~\ref{FIG:neg_mixed}c).

Finally, we also performed the same experiment with the weak-field light source (see Supplementary Information). Similar to the results obtained for the single photon sources, we also observed negative values. In this experiment, we post-selected the raw data to evaluate the negativity in a way that only single APD clicks were sampled and the rest of events, e.g., more than two clicks simultaneously were neglected. In general, the weak-field light is understood not to be the single-photon source in the sense that this light does not show the anti-bunching effects. However, the negativity can be detected with a post-selection process. Recently, such a phenomenon was reported that a coherent state of the optical field can show the nonclassicality~\cite{Jae17, Zhang19}. We highlight that the operational quasiprobability reveals the negativity by an interplay between given state and measurement.

In conclusion, we experimentally explored the negativity of the operational quasiprobability by measuring the single photon polarizations. We introduced the context of measurement selection by constructing the quasiprobability such that by marginals it provides the same fundamental entity as that of the single measurements. As a result, the quasiprobability can reproduce the quantum predictions by allowing negative probabilities. The measured negatives highlight the discrepancy between the classical and quantum predictions in the context of measurement selection. In the case of the classical prediction, we investigated the macrorealistic model assuming the NSIT and AoT conditions. In this model, the operational quasiprobability becomes a legitimate joint probability distribution for the given measurement setups. Therefore, observing the negatives highlights the nonclassical property in the context of measurement selection. We note that negativity is merely a sufficient condition for violating the NSIT. That is, the operational quasiprobability can be nonnegative even if the NSIT is violated. Such a case is encountered if a general measurement is involved, which will be discussed in a forthcoming paper. From a fundamental perspective, the measured negativity provides an operational approach to unravel the nonclassicality of photons in the context of measurement selection.




\section*{Methods}

{\bf Input preparation.} 
Heralded single photons: The experimental schematic used to generate the heralded single photons is shown in Fig.~\ref{FIG:exp_scheme}b. Orthogonally polarized photon pairs are generated by a type-II SPDC process using a 401.5 nm continuous wave (CW) mode laser to irradiate a periodically poled KTiOPO$_4$ (PPKTP). 
The resulting photon pair is separated using a PBS. The horizontally polarized photon (signal) is used as the input photon and the vertically polarized photon (idler) is used as the trigger. Simultaneous detection in the signal and idler channels exclude the contribution of the vacuum state of the SPDC source, and thus gives the anti-bunching property of the heralded single photons. We also experimentally examined this by measuring a second-order correlation function with multi-channel correlation measurements and obtained a value of $g^{2}(0)=0.036$ (for more details, see Supplementary Information).


Single photons from a single molecule: The output of a CW-mode laser (532 nm) is focused using an oil objective (NA=1.40) onto a single terrylene molecule which is embedded in a thin para-terphenyl crystal ($\sim$20 nm) in a total internal reflection geometry~\cite{PFAB04}. The emitted fluorescence signal transmitted across a long-pass filter (LPF) is collected by the same objective and diverted to the detection part.
The measurement value of $g^{2}(0)$ is 0.14 (see Supplementary Information).

Input polarization state: The input photon polarization state was set to a pure state for H/V polarization states in the form of $\ket{\Psi ( \theta, \phi)} = \cos (\theta/2) \ket{H} + e^{i\phi} \sin (\theta/2)\ket{V}$, where $\theta$ and $\phi$ are the bases of the polar coordinates on a unit sphere, called a Bloch sphere. To prepare such a state, the horizontally polarized photons are sequentially transmitted through the half wave-plate (HWP) and two quarter wave plates (QWPs) as illustrated in Fig.~\ref{FIG:exp_scheme}b. The last QWP is fixed at the angle of $\pi/4$. The final polarization state of the input photon is obtained as follows:
\begin{eqnarray}
&&T_{\rm{QWP}} \left(\frac{\pi}{4}\right)  T_{\rm{HWP}} (p) T_{\rm{QWP}} (q)
\begin{pmatrix}
1\\
0
\end{pmatrix} \nonumber \\
&=& e^{i (-2p+q+\pi/4)}
\begin{pmatrix}
\cos (\pi/4-q)\\
e^{i (4p-2q-\pi/2)} \sin(\pi/4-q)
\end{pmatrix},
\end{eqnarray}
where $T_{\rm{Q(H)WP}}$ represents the transfer matrix of the corresponding wave-plate. The parameters of $p, q$, and $\pi/4$ are the rotating angles of the waveplates and follow the relations: $p=(\pi + \phi -\theta)/4$ and $q=(\pi/2-\theta)/2$.

 

{\bf Collecting data.} In the detection part, we counted the relative ratio of measurement outcomes as follows: $P_{n_1,n_2}(a_1,a_2)=N_{n_1,n_2}(a_1,a_2)/N^{\textrm{tot}}_{n_1,n_2}$, where $N_{n_1,n_2}(a_1,a_2)$ denotes the sum of the counted photons at the detector $D_{a_1,a_2}$ and $N^{\textrm{tot}}_{n_1,n_2}$ is the total number of counted photons at all detectors for a given setup $(n_1,n_2)$. We here describe in detail how to obtain each component in Eq.~(\ref{eq:w}) in terms of the measurement setting $(n_1, n_2)$ and the detectors (APDs) $D_{a_1, a_2}$.

\begin{enumerate}

\item $P_{t_1,t_2}(a_1,a_2)$: we implement the consecutive measurement of H/V and D/A, which corresponds to $P_{n_1, n_2} (a_1, a_2)$  with $(n_1, n_2)=(1,1)$. Then, collect the data at all the APDs $D_{a_1, a_2}$.

\item $P_{t_1}(a_1)$: this corresponds to the H/V single measurement that can be rewritten by $P_{n_1, n_2}(a_1,0)$ with $(n_1, n_2)=(1,0)$. In laboratory, we fixed the PBS$_{2,3}$, thus it can be obtained by the marginal of the $P_{n_1,n_2}(a_1,a_2)$ with $(n_1, n_2)=(1,1)$. That is, $P_{1,0}(a_1,0)=P_{1,1}(a_1,0)+P_{1,1}(a_1,1)$. In this case, the data is collected at $D_{a_1, 0}$ for $P_{1,1}(a_1,0)$ and $D_{a_1, 1}$ for $P_{1,1}(a_1,1)$.

\item $P_{t_1,t_2}(a_1)$: this is simply a marginal of $a_2$ by $P_{t_1,t_2}(a_1,a_2)$.

\item $P_{t_2}(a_2)$: this corresponds to the D/A single measurement that is given by $P_{n_1, n_2}(0, a_2)$ with $(n_1, n_2)=(0,1)$. After the PBS$_{1}$ is out, we collect the data at APDs $D_{0,a_2}$.

\item $P_{t_1,t_2}(a_2)$: this is simply a marginal of $a_1$ by $P_{t_1,t_2}(a_1,a_2)$.

\end{enumerate}

{\bf Detection and data acquisition.}
The photon clicks of the single photon counting detectors (Perkin-Elmer, SPCM-AQ4C) are sent to a field-programmable gate array (FPGA, NI PXI-7841R) for the post-selection process. For data acquisition and processing, the FPGA operates at a 25 ns clock speed (40 MHz) and 125 ns processing cycle. Each detector has a slightly different detection efficiency; therefore, it is of crucial importance that their effective efficiencies are equalized. To this end, we measured the counts of the input polarization with the angle
$\theta=45 \degree$, where all four detectors are supposed to have the same input photon flux under the same detection efficiency. The detected counts are used as the references to normalize the signals of each detector. 

The sum of the detected counts at four APDs is about $10^4$ on average, for all values of $\theta$ and $\phi$. Table \ref{table1} shows some examples of the detected counts for each APD from Fig.~\ref{FIG:neg_spdc}a.

\begin{table}[h]
\caption{\label{table:xxx} Some examples of the detected counts for each APD. The data is from Fig.~\ref{FIG:neg_spdc}a, i.e., the angle $\theta$ with $\phi=0^{\circ}$.}
\begin{ruledtabular}
\begin{tabular}{cccccc}
$\theta (\degree)$ & PBS1 on/off & $D_{0,0}$ & $D_{0,1}$ & $D_{1,0}$ & $D_{1,1}$ \\ 
\colrule
 0 & on &  4955 & 5018 & 16 & 11 \\
  & off &  5058 & 4940 & 2 & 0 \\ 
 
 45 & on & 4152 & 4262 & 791 & 795 \\
  & off & 8470 & 1529 & 0 & 1 \\
 
 90 & on & 2430 & 2593 & 2411 & 2567 \\ 
  & off & 9972 & 28 & 0 & 0 \\ 
\end{tabular}
\end{ruledtabular}
\label{table1}
\end{table}


\section{Supplementary Information}

\subsection{Error analysis}
\label{SEC:ERROR}

\begin{figure*}[t]
	\centering
	\includegraphics[width=1\textwidth]{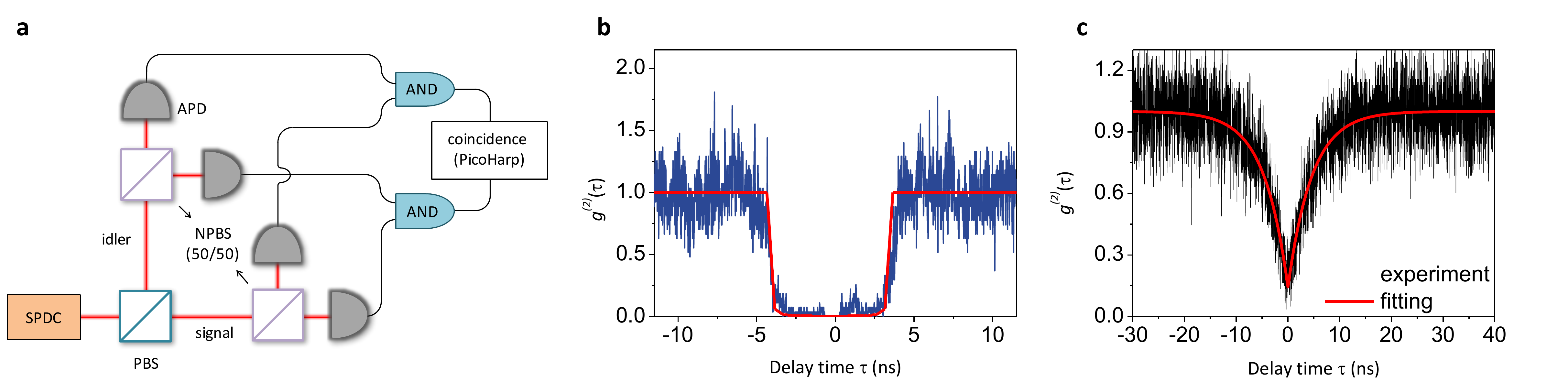}
	\caption{{\bf Second order correlation function of the resources.} {\bf a}, Experimental schematic for measuring the anti-bunching property of heralded single photons~\cite{RAZAVI09, Bashkansky14}. {\bf b}, Experimental (blue line) and theoretical values (red line) of the second order correlation function of heralded single photons. Average value of $g^{(2)}$ dip is 0.036. {\bf c}, The $g^{(2)}(\tau)$ function of single photons emitted from a single terrylene molecule. The result reads $g^{(2)}(0)=0.14$.}
	\label{FIG:g2}
\end{figure*}

Experimental errors occur when optical devices (components) have imperfect alignment and/or response. We characterized the optical components used in our measurements to obtain their operational errors; PBSs, wave-plates, avalanche photodiodes (APDs, Perkin Elmer, SPCM AQ4C). For example, the angles of the wave-plate are controlled by motorized rotational stages, and the operating error is less than 0.5 degree. This results in approximately 1 degree of maximum error for the input polarization, which leads to $1.1\%$ error in the detected counts at the APDs. Also, the extinction ratio of the PBS (CVI Laser Optics, PBS-800-050) exceeds $10^3$. This is true in the transmitted part ($<0.1\%$ error), but the reflected beam contains about $5\%$ of incorrectly polarized photons. The four APDs are corrected using the reference detection efficiency as described in the detection scheme in the Method section. However, this cannot be perfectly accomplished. Thus, we consider the $5\%$ error including the fluctuation of the APD’s working efficiency. The total error in the measurement of a quantity is calculated as $\sum_{i} \sqrt{\text{(error of each component}_{i})^2 \times \text{(times used)}}$. Here, we assumed that there is no correlation between the errors of different components. Note that the given error values (error bars in the figures) are maximally estimated. The statistical fluctuation over measurements at many time intervals can be inferred from the distribution of the experimental values (red circles) in Fig.~2a in the main text, which is much smaller than the given error bar.

\begin{figure*}[t]
	\centering
	\includegraphics[width=0.95\textwidth]{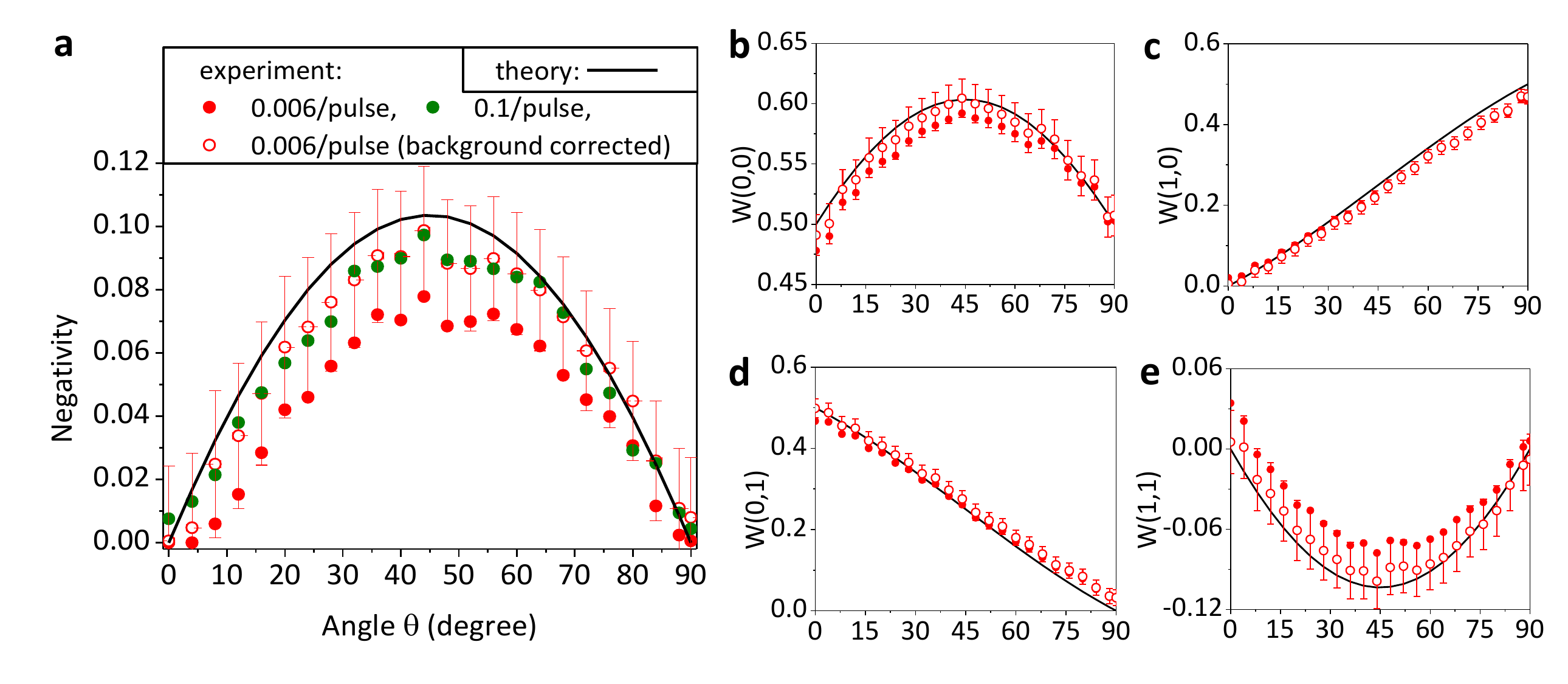}
	\caption{{\bf Negativity for a weak field input.} {\bf a}, Negativity for two different input intensities (average photon number): Green circle for $1 \times 10^{-1}$ per pulse, and red circle for $6 \times 10^{-3}$ per pulse. When the dark-count of the APDs is corrected, the experimental results are in closer agreement with the theoretical data as shown by the red open circles. The error bars are shown only for the red open circles for clarity. Each component of the quasiprobability distribution $\mathcal{W}(a_1,a_2)$ is presented in {\bf b}-{\bf e} as a function of the angle $\theta$ with a fixed $\phi = 0^{\circ}$ of the state $\ket{\Psi ( \theta, \phi)} = \cos (\theta/2) \ket{H} + e^{i\phi} \sin (\theta/2)\ket{V}$.}
	\label{FIG:weak}
\end{figure*}

\subsection{Second order correlation function of the resources}
\label{APX:APXB}
We examined the anti-bunching characteristics of the heralded single photons and single photons from a single molecule by measuring the second-order correlation function $g^{(2)}(\tau)$. In the case of the SPDC source, as shown in Fig.~\ref{FIG:g2}a, photon pair is initially separated using a PBS into the signal and the idler paths. Both paths are then sub-divided by the 50:50 non-polarizing beam splitter (NPBS) into two branches. By combining one branch of the signal and one of the idler using an AND gate, it becomes possible to mimic the coincidence of the heralded photon. Therefore, the Hanbury-Brown and Twiss (HBT) measurement of the outputs of two AND gates implies the $g^{(2)}(\tau)$ of the heralded photons. Two digital chips (PO74G08) are used for the AND gates. The HBT measurement was performed using the start-stop mode of a TCSPC device (PicoQuant, PicoHarp 300). 
The resulting experimental curve is shown together with a theoretical calculation in Fig.~\ref{FIG:g2}b. In the calculation, we considered the following system parameters; the timing jitter of the APDs ($= 0.61$ ns) and the coincidence time window of the AND gates ($= 5.5$ ns). The average value of the dip in the time delay range of -3 ns $\sim$ 3 ns is only 0.036. The near zero value of $g^{(2)}(0)$ ensures that the heralded photons are the most similar to the single photons.
For the case of the single molecule, the emitted photons are divided by a NPBS into two branches and are directly used as the inputs of the start-stop measurements. The results are shown in Fig.~\ref{FIG:g2}c.

\subsection{Additional experiment by post-selected weak field}
\label{APX:weak}

We here discuss the negativity of the operational quasiprobability using a weak-field. Given that such light does not exhibit the anti-bunching characteristic, the weak-field can be regarded as classical light. However, we can detect the negativity with a post-selection process. This indicates that our method provides an operational way to detect the nonclassicality of optical fields within the context of selecting measurement procedures.

The input source was prepared as follows; we used picosecond pulses from a mode-locked Ti:sapphire laser (Mira 900). The centre wavelength is set to 800 nm and the pulse repetition rate is reduced down to 3.8 MHz with a pulse picker (Coherent 9200). Then, using neutral density filters, the intensity of the beam is attenuated so that the average number of photons range from $10^{-3}$ to $10^{-1}$ per pulse.
We implemented the same measurement setups as shown in Fig. 1 in the main text. We post-selected the raw data  to evaluate the negativity in a way that only single APD clicks were sampled and the rest of events, e.g., more than two clicks simultaneously were neglected.

Experimental results together with the theoretical predictions are presented in Fig.~\ref{FIG:weak}a for the negativity and in Fig.~\ref{FIG:weak}b-e for each $\mathcal{W}(a_1, a_2)$. We started with the average photon number $6 \times 10^{-3}$ per pulse (red open and filled circles in Fig.~\ref{FIG:weak}a). The maximum negativity was obtained as $0.078$ without the correction of the dark count of the APDs (red filled circle). This maximum increased to $0.099$ after correcting for the dark count (red open circles); we measured the dark count of each APD and subtracted this value from the total measured counts. For a higher average photon number ($10^{-1}$ per pulse), the maximum negativity was $0.097$ even without the correction for dark-count (see green circle). This is because for a higher detection count, the contribution of the dark count of the APD ($\sim10^3$ counts per second) becomes smaller. Note that all maxima are obtained for $\theta=44\degree$ and $\phi=0^{\circ}$. We followed the error analysis in Sec.~\ref{SEC:ERROR}.

\section{acknowledgements}
This research was supported by the National Research Foundation of Korea (NRF) grant (No. NRF-2019R1A2C2005504 and No. 2016R1A2B4014370), funded by the MSIP (Ministry of Science, ICT and Future Planning), the Korean government and was also supported by the MSIT (Ministry of Science and ICT), Korea, under the ITRC (Information Technology Research Center) support program (IITP-2018-2015-0-00385) supervised by the IITP (Institute for Information \& communications Technology Promotion). JR acknowledges the National Research Foundation, the Prime Minister’s Office, Singapore and the Ministry of Education, Singapore under the Research Centres of Excellence programme and Singapore Ministry of Education Academic Research Fund Tier 3 (Grant No. MOE2012-T3-1-009).

\section{Data Availability}
All relevant data are available from the corresponding author on reasonable request.

\section{Author contributions}
S.H., J.-S.L., K.H.S., Jiwon L. and K.-G.L. devised and performed the experiment; J.R., J.J., James L. and Jinhyoung L. developed the theoretical tools; all authors discussed the results and contributed to the writing of the manuscript.

\section{Competing interests}
The authors declare that there are no competing interests.

\end{document}